\documentclass[11pt]{article}
\usepackage{graphicx}
\def\bmath#1{\mbox{\boldmath$#1$}}
\newcommand{\huts}{Hutsem\'{e}kers }
\title{Large Scale Alignment of Optical Polarizations from Distant QSOs
using Coordinate Invariant Statistics}
\author{Pankaj Jain, Gaurav Narain and S. Sarala\\
Physics Department, I.I.T. Kanpur, India 208016 \\
email : pkjain@iitk.ac.in}
\begin{document}
\maketitle

\begin{abstract}
We introduce several coordinate invariant statistical procedures
in order to test for local alignment of polarizations. A large
scale alignment of optical polarizations from 
distant QSOs has recently been observed by \huts and 
collaborators. The new statistical procedures are based on comparing
polarizations at different angular coordinates by making a  parallel 
transport. The results of these statistical procedures continue
to support the existence of the large scale alignment effect
in the QSO optical polarization data. 
The alignment is found to be much more pronounced in the data sample
with low degrees of polarization $p\le 2$\%. This suggests that the
alignment may be attributed to some propagation effect. 
The distance scale over which the alignment effect is dominant is 
found to be of order 1 Gpc. 
We also find
that a very large scale alignment is present in the large redshift, 
$z\ge 1$, data sample. Infact the data sample with 
$z\ge 1$ appears to be aligned over the entire celestial sphere.  
We discuss possible physical effects, such 
as extinction and pseudoscalar-photon mixing, which may be responsible
for the observations. 

\end{abstract}

\section{Introduction}
The optical polarizations from QSOs
appear to be aligned with one another over very large 
distances (\huts 1998). This effect was further confirmed 
by \huts \& Lamy (2001) 
using a larger data set.
It was observed that 
the polarizations from QSOs in any particular
spatial region or a patch have a tendency to align with one another, 
without any evidence of a large
scale anisotropy. 
The effect was found to be redshift dependent, namely the patches
which are aligned are delimited both in angular as well as radial (redshift)
coordinates. A very striking alignment was found in the region, 
called A1 by 
\huts (1998), delimited in Right Ascension by $11^{\rm h}15^{\rm m}
\le {\rm RA} \le 14^{\rm h}29^{\rm m}$ and in redshift by $1.0\le z\le 2.3$.
By using several statistical tests (\huts 1998; \huts \& Lamy 2001) the authors
were able to rule out the hypothesis of uniform distribution 
at approximately 0.1\%
significance level. The basic idea behind these statistical procedures
is to test for the dispersion in the orientations of polarizations
in any small neighbourhood. The tests require comparing the polarization
vectors located at different angular coordinates. These 
polarizations are specified by the angles they make with respect
to the local meridians (\huts 1998). The statistical tests used 
by \huts (1998) are, however, dependent on the precise position
of the pole used for defining the coordinate system and hence 
it is difficult to interpret the results obtained by these tests.
In the present paper we introduce
some statistical tests  which are independent of the
coordinate system and apply them to the
QSO optical polarization data.

\section{Parallel Transport}
The basic problem can be explained as follows. 
Consider two tangent vectors $\bmath{v_1}$ and $\bmath{v_2}$ located
at two different points P$_1$ and P$_2$ on the surface of the 
sphere. 
These vectors are specified at each point on the sphere 
in terms of the angle they make 
with respect to the local unit vector $\hat\phi$. Here we have
chosen the spherical polar coordinate system defined by the unit
vectors $(\hat r,\hat\theta,\hat \phi) $. 
Let $\alpha_1$ be the
angle made by the vector $\bmath{v_1}$ 
with respect to the local unit vector $\hat\phi_1$ 
at the location P$_1$. Similarly let $\alpha_2$ be the 
angle made by the vector $\bmath{v_2}$ with respect to the local unit 
vector $\hat\phi_2$ at the location P$_2$. 
The two vectors in their local coordinates are given by
\begin{eqnarray}
\nonumber
\bmath{v_1} = \cos\alpha_1\hat\phi_1 + \sin\alpha_1\hat\theta_1
\\ 
\bmath{v_2} = \cos\alpha_2\hat\phi_2 + \sin\alpha_2\hat\theta_2
\nonumber
\end{eqnarray}
In order to compare these two vectors we may use the usual dot product,
\begin{equation}
\bmath{v_1\cdot v_2} = \cos(\alpha_1)\cos(\alpha_2)+
\sin(\alpha_1) \sin(\alpha_2)\ .
\end{equation}
However this dot product is not invariant under coordinate transformation
and hence a statistical procedure based on this will be coordinate
dependent. A coordinate invariant dot product may be obtained if we 
parallel transport $\bmath{v_1}$ to P$_2$ and then take the
dot product of the transported vector $\bmath{v_1}'$ 
with $\bmath{v_2}$. Let the parallel 
transported vector $\bmath{v_1}'$ make the angle $\alpha_1'$ 
with respect to $\hat\phi_2$. We are interested in finding $\alpha_1'$
which can then allow us to calculate the angle $\alpha_1' -
\alpha_2$ between the parallel transported vector 
$\bmath{v_1}'$ and the vector $\bmath{v_2}$.

Let $\hat r_1$ and $\hat r_2$ be unit radial vectors
at the points P$_1$ and P$_2$ respectively. The unit vector $\hat s$ 
perpendicular
to the plane containing these two radial vectors is given by
\begin{equation}
\hat s = {\hat r_1\times \hat r_2\over |\hat r_1\times \hat r_2|}
\end{equation}
Therefore the unit tangent vectors $\hat t_1$ and $\hat t_2$ at 
the two points P$_1$ and P$_2$ respectively along the
great circle passing through these two points are given by,
\begin{equation}
\hat t_1 = \hat s\times \hat r_1
\end{equation}
\begin{equation}
\hat t_2 = \hat s\times \hat r_2
\end{equation}
In terms of the local basis $(\hat \theta,\hat\phi)$ these two vectors
are given by
\begin{equation}
\hat t_1 = \hat \theta_1\cdot \hat t_1 \hat \theta_1 + \hat\phi_1\cdot \hat t_1 \hat \phi_1
\end{equation}
\begin{equation}
\hat t_2 = \hat \theta_2\cdot \hat t_2 \hat \theta_2 + \hat\phi_2\cdot \hat t_2 \hat \phi_2
\end{equation}
where 
\begin{equation}
\hat \theta_1\cdot \hat t_1 = {-\sin\theta_1 \cos \theta_2 + \cos\theta_1
\sin\theta_2\cos(\phi_2-\phi_1)\over \sqrt{1-(\hat r_1\cdot\hat r_2)^2}}
\end{equation}
\begin{equation}
\hat \phi_1\cdot \hat t_1 = {\sin\theta_2 
\sin(\phi_2-\phi_1)\over \sqrt{1-(\hat r_1\cdot\hat r_2)^2}}
\end{equation}
\begin{equation}
\hat \theta_2\cdot \hat t_2 = -{-\sin\theta_2 \cos \theta_1 + \cos\theta_2
\sin\theta_1\cos(\phi_2-\phi_1)\over \sqrt{1-(\hat r_1\cdot\hat r_2)^2}}
\end{equation}
\begin{equation}
\hat \phi_2\cdot \hat t_2 = {-\sin\theta_1 
\sin(\phi_1-\phi_2)\over \sqrt{1-(\hat r_1\cdot\hat r_2)^2}}
\end{equation}

\begin{figure}[]
\centering
\includegraphics[]{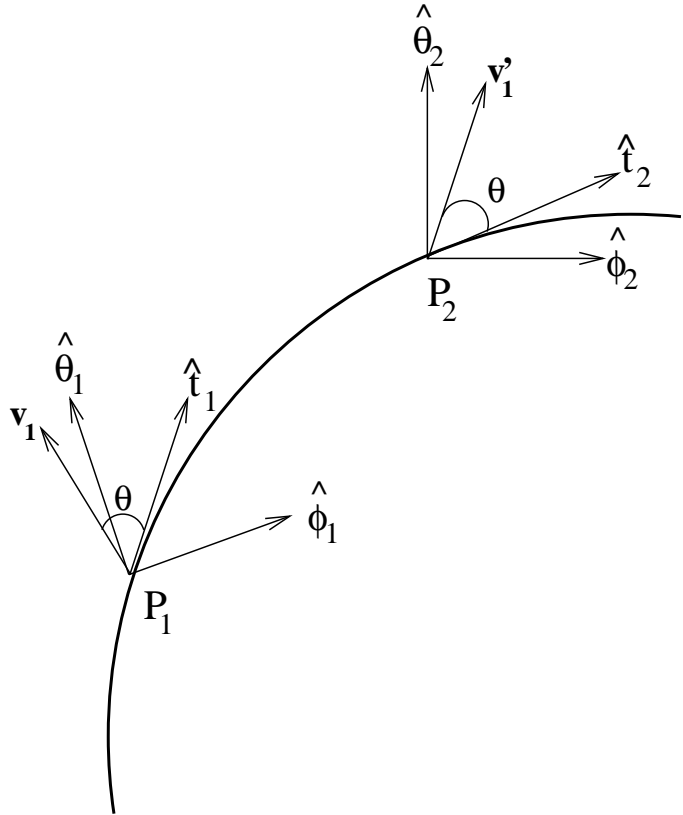}
\caption{The curve is the great circle passing
through points $P_1$ and $P_2$. $t_1$ and $t_2$ are the tangent
vectors at points $P_1$ and $P_2$ respectively. Polarization
vector $v_1$ from location $P_1$ is parallel transported to 
location $P_2$. Parallel transported vector is indicated by $v_1'$.}
\label{transport}
\end{figure}

As we parallel transport a vector along the great circle its angle
with respect to the tangent to the great circle remains fixed. Hence
in order to determine the angle by which the vector has turned due
to parallel transport we only need to find the 
orientation of $\hat t_1$ and $\hat t_2$ with respect to the
local basis at the points $P_1$ and $P_2$ respectively.
The angle $\xi_1$ between $\hat t_1$ and 
$\hat \phi_1$ is given by $\cos^{-1}(\hat \phi_1\cdot \hat t_1)$. 
This gives $\xi_1$ upto an overall addition of $\pi$, 
which is fixed by the sign of
$\hat \theta_1\cdot \hat t_1$. 
If $\hat \theta_1\cdot \hat t_1<0$ then $\xi_1$ 
lies in the third or fourth quadrand. The angle $\xi_2$
is obtained in a similar manner. Hence the transported vector 
$\bmath{v_1}'$ makes an angle $(\alpha_1+\xi_2-\xi_1)$ with respect 
to the $\hat \phi_2$. Therefore the rule for comparing vectors at two 
different locations on the sphere is given by the generalized dot 
product,
\begin{equation}
\bmath{v_1}\odot\bmath{v_2} = |\bmath{v_1}| |\bmath{v_2}| \cos(\alpha_1-
\alpha_2+\xi_2-\xi_1)\ ,
\label{GDP1}
\end{equation}
i.e. the dotproduct
between the parallel transported vector $\bmath{v_1}'$ and the
vector $\bmath{v_2}$.

In our discussion so far we have explained the procedure for comparing
the vectors at two different points. The polarization
orientations $\alpha_1$ and $\alpha_2$
at two different points $P_1$ and $P_2$ respectively can also 
be compared in a similar manner. 
We associate the unit vectors $\bmath{v_1}=[\cos(2\alpha_1),\sin(2\alpha_1)]$
and $\bmath{v_2}=[\cos(2\alpha_2),\sin(2\alpha_2)]$ with the polarizations
at $P_1$ and $P_2$ respectively.
We parallel transport the
polarization $\alpha_1$ to the location $P_2$. The angle between
the parallel transported polarization and the unit vector
$\hat\phi_2$ along the local $\phi$ axis at $P_2$ is equal to 
$\alpha_1^\prime = (\alpha_1+\xi_2-\xi_1)$. We can assign a 
unit vector $\bmath{v_1^\prime}=
[\cos(2\alpha_1^\prime),\sin(2\alpha_1^\prime)]$ with this 
transported polarization.
We then define the generalized product $(\alpha_1,\alpha_2)$ between
the two polarizations as the ordinary
dot product between the two vector $\bmath{v_1^\prime}$ and $\bmath{v_2}$
\begin{equation}
(\alpha_1,\alpha_2) = \bmath{v_1^\prime}\cdot\bmath{v_2} = \cos[2(\alpha_1-
\alpha_2+\xi_2-\xi_1)]\ .
\label{GDP}
\end{equation}

\section{Statistical Tests for Alignment}
The simplest statistics to test the alignment of vectors is to
calculate the dispersion of the polarization with respect
to its nearest neighbours. Consider the $n_v$ nearest neighbours of the
vector $\bmath{v_i} = [\cos(2\theta_i), \sin (2\theta_i)]$ located at the
i-th location, including the vector $\bmath{v_i}$ itself. 
Here $\theta_i$ is the polarization orientation. The orientation angle may
be measured with respect to the local meridian or the local latitude. 
Our final results depend only on the differences of angles and hence
our formulae are directly applicable for either definition.
The nearest neighbours are determined by computing the relative
distance in three dimensions with the co-moving radial distance $r(z)$ given 
in terms of the redshift $z$ by the standard relation,
\begin{equation}
r(z) = \frac{2c}{H_0}\left[1 - (1+z)^{-1/2}\right]
\label{dis}
\end{equation}
where $H_0$ is the Hubble constant and $c$ is the velocity of light.
A measure of the dispersion 
$d_i$ of these set of vectors is given by,
\begin{eqnarray}
\nonumber
d_i(\theta) & = & \sum_{k=1}^{n_v} (\theta,\theta_k)\\
& = & \sum_{k=1}^{n_v} \cos\left[2\theta - 2\left(
\theta_k + \Delta_{k\rightarrow i}\right) \right]
\label{dispersion}
\end{eqnarray}
where $\Delta_{k\rightarrow i}$ is the angle by which the polarization 
orientation angle $\theta_k$ changes after being parallel 
transported to the position of the polarization angle $\theta_i$
and $\theta$ is a measure of the mean polarization at the 
i-th position. 
The symbol $(\theta,\theta_k)$ is defined in Eq. \ref{GDP}.
Hence $d_i$ is calculated by parallel transporting
the $n_v$ nearest neighbours of the polarization $\theta_i$
to the i-th position and
then taking the dot product of the resulting $n_v$ vectors
$\bmath{v_k^\prime} = [\cos(2\theta_k + 2\Delta_{k\rightarrow i})]$
with the vector
$\bmath{v}(\theta) =[\cos(2\theta),\sin(2\theta)]$. 
The $n_v$ nearest neighbours include the vector $\bmath{v_i}$ itself.
The magnitude 
of $d_i(\theta)$ is then maximized with respect to $\theta$. The 
resulting vector $\bmath{v}(\theta)$ gives a measure of the local
mean direction and the magnitude of the maximized $d_i(\theta)$ gives 
the measure of dispersion.  We define the statistic as follows, 
\begin{equation}
S_D^p = \frac{1}{n}\sum_{i=1}^{n} d_i{\big |}_{\rm max}
\end{equation}
where the sum is over the entire data sample.
A large value of $d_i$ implies small dispersion and hence
 a large $S_D^p$ would imply a strong 
alignment between the polarization vectors.

\begin{figure}[]
\centering
\includegraphics[]{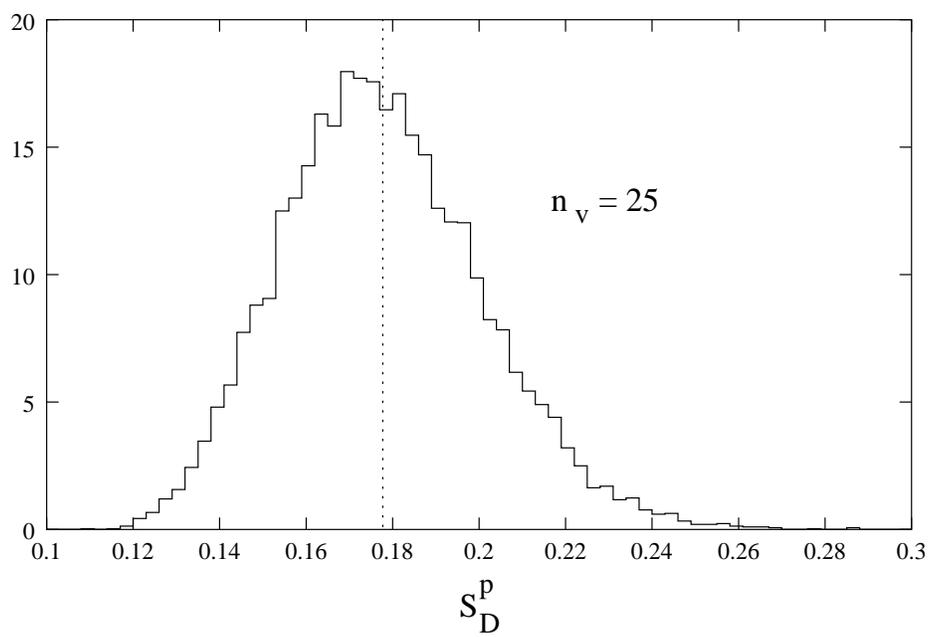}
\caption{The histogram of the statistic $S_D^p$ using 10000 random sample
with the number of nearest neighbours $n_v=25$. The mean position
of the statistic for a random sample is shown by the verticle line.} 
\label{histogram}
\end{figure}

It is clear that
the mean value $\overline{S_D^p}$ of this statistic for a random sample 
is proportional
to $1/\sqrt{n_v}$. We verify this explicitly by numerical simulations. For
$n_v= 25$ we find that the mean value $\overline{S_D^p}=0.178$.
In Fig. \ref{histogram} we show the histogram of the statistic 
$S_D^p$ obtained by using 10000 random samples with $n_v= 25$. 
The distribution tends to a normal distribution for large data 
samples.
The distribution for different values of $n_v$ was found to be 
identical except for a shift in the mean position, which as mentioned
above, varies as $1/\sqrt{n_v}$. In our
calculations we evaluated the statistical significance level (S. L.) for 
each case by explicit numerical simulation using a large number of 
random samples.
The S. L. is defined as the probability that the value of the statistic
obtained from the data may be obtained as a statistical fluctuation
from a random sample. 
It is numerically evaluated by computing the statistic $(S*)$ of the
data sample with the statistic $S$  obtained from a large number of random 
samples. The S. L. is a probability that $S<S*$.

Our second statistical test is the modified Andrews \& 
Wasserman test, which was used by \huts (1998) and \huts \& Lamy (2001).
We first review this test. For each vector 
$\bmath{v_j} = (\cos 2\theta_j, \sin 2\theta_j)$
the mean resultant vector 
with respect to its $n_v$ nearest neighbours 
can be written as 
\begin{equation}
\bmath{V_j} = \frac{{\cal N}}{n_v}\left(\sum_{k=1}^{n_v} 
\cos 2\theta_k, \sum_{k=1}^{n_v} \sin 2\theta_k \right) \ .
\label{AW1}
\end{equation}
where ${\cal N}$ is a normalization constant such that the vector $V_j$
is normalized to unity.
The sum over the nearest neighbours $n_v$ excludes the $j$-th vector itself. 
We then define the mean direction ${\bar{\theta_j}}$ such
that $\bmath{{V_j}} = (\cos 2\bar{\theta_j},\sin 2\bar{\theta_j})$. 
A measure of the alignment of vector 
$\bmath{v_i} = (\cos 2\theta_i, \sin 2\theta_i)$
with the $n_v$ nearest neighbours of the vector 
$\bmath{v_j} = (\cos 2\theta_j, \sin 2\theta_j)$
is given by the dot product
$D_{ij} = \bmath{v_i} \cdot \bmath{V_j}$.
In order to evaluate the statistic, the $D_{ij}$, $j=1,..,n$, values are
sorted in the ascending order for each $i$. 
The rank ${\cal R}_i$ of $D_{i,j=i}$
is then evaluated. If $n$ is the
number of sources in the sample then the statistic is given as
\begin{equation}
Z_c = \frac{1}{n_v}\sum_{i=1}^{n} \frac{{\cal R}_i - (n+1)/2}{\sqrt{n/12}}
\label{Zc}
\end{equation}
As shown by Bietenholz (1986), the statistic $Z_c$ approximately follows 
a normal distribution.
However we did not use this asymptotic result in obtaining our 
S. L. but evaluated it explicitly by using a large number of random samples.
The Andrews and Wasserman test can be modified (\huts 1998) by
using the unnormalized vector $V_j$, i.e. by setting the normalization
constant ${\cal N}$ to unity. The resulting statistic is denoted
by $Z^m_c$ (\huts 1998). 

As explained earlier this statistic is not coordinate invariant since 
it compares the vectors at different locations without making a parallel
transport.
We can modify this statistic by using the generalized product,
Eq. \ref{GDP}, to compare polarizations at different positions. We therefore
express $D_{ij}$ directly in terms of the coordinate invariant quantity 
$(\theta_i,\theta_k)$. The basic idea behind the calculation of
$D_{ij}$ is to compare i-th polarization $\theta_i$ with the 
$n_v$ nearest neighbours $\theta_k$ of the j-th polarization $\theta_j$.  
Hence one possible coordinate
invariant definition of $D_{ij}$ is
\begin{eqnarray}
\nonumber
D_{ij}^p & = & \frac{1}{n_v}\sum_{k=1}^{n_v} (\theta_i,\theta_k) \\
& = & \frac{1}{n_v} \sum_{k=1}^{n_v} \cos\left[2\theta_i - 2\left(
\theta_k + \Delta_{k\rightarrow i}\right)
\right]
\label{dmatrix}
\end{eqnarray}
where $\theta_k$, $k=1,...,n_v$ are the $n_v$ nearest neighbours of 
the polarization 
$\theta_j$, excluding $\theta_j$ itself. 
Hence all the $n_v$ polarizations are compared to the polarization 
$\theta_i$ after 
parallel transporting them to the i-th position.
It is clear that if we replace the
generalized product with the ordinary dot product in this equation, 
i.e. if we set $\Delta_{k\rightarrow i}=0$, then $D_{ij}^p$
reduces to the matrix $D_{ij}$ that is used in the Andrews \& 
Wasserman test with the normalization factor ${\cal N}$ in Eq. \ref{AW1}
set equal to unity.
$D_{ij}^p$ is the measure of alignment
between the i-th vector and $n_v$ nearest neighbours of the j-th vector.
The statistic is then calculated in a manner similar to that used
in the Andrews \& Wasserman test. We again arrange $D_{ij}^p$, $j=1,..,n$,
in ascending order and obtain the rank ${\cal R}_i$ of 
$D_{i,j=i}^p$. The statistic is calculated using equation
\ref{Zc} and is called $Z_c^p$.

Another somewhat more complicated coordinate invariant definition 
of $D_{ij}$ can be obtained by parallel transporting all the polarizations 
to the position of the polarization $\bmath{v_j}$ and then comparing them at
this point. The resulting expression is given by,
\begin{equation}
\nonumber
D_{ij}^{p\prime} =  \frac{1}{n_v} \sum_{k=1}^{n_v} \cos\left\{ 
\left[2\theta_j - 2\left(\theta_k + \Delta_{k\rightarrow j}\right) 
\right] - \left[2\theta_j - 2\left(\theta_i + \Delta_{i\rightarrow j}
\right) \right] \right\}
\label{dmatrixprime}
\end{equation}
This can also be explicitly expressed in terms of the basic 
coordinate invariant quantity, 
$(\theta_i,\theta_j)$, but the precise expression 
is not needed here.
The results obtained by using $D_{ij}^{p\prime}$ are similar to
those obtained by using $D_{ij}^{p}$.

\subsection{Coordinate Invariance}
We next test explicitly whether the statistics defined in the 
previous section are indeed invariant under coordinate transformations.
Consider a polarization which is oriented
at an angle $\theta$ and 
located at the position ($\alpha,\delta$) 
in a particular coordinate system. 
Next we consider a transformed coordinate frame such that its pole
lies at the position ($\alpha_p,\delta_p$) in the original coordinates. 
Let the polarization be oriented at an angle $\theta_N$ in the new
system. The angle
$\theta_N$ can be expressed as (\huts 1998)
\begin{equation}
\tan\left( \theta - \theta_N \right) = \frac{\cos\delta_p
\sin(\alpha_p - \alpha)}{\sin\delta_p\cos\delta -
\sin\delta\cos\delta_p\cos(\alpha_p - \alpha)}\ .
\label{transformation}
\end{equation}
The angular position of the polarization can also be evaluated
in the new coordinate system by making an orthogonal transformation.
The coordinates are transformed by first making a rotation about 
x-axis by  an angle $( - \delta_p + \pi/2)$ and then a rotation about 
the new z-axis by an angle $(-\alpha_p + \pi/2)$.
The resulting statistics $S_D^p$ and $Z_c^p$ for several 
choices of the pole position are given in Table 1, choosing the
number of nearest neighbours $n_v=20$. The coordinate independence
of the statistic is clear from this table.

\begin{table}[h!]
\begin{center}
\begin{tabular}{|c|c|c|c|}\hline \hline
$\bmath{\alpha_p}$ & $\bmath{\delta_p}$ & 
$\bmath{S_D^p}$ & $\bmath{Z_c^p}$\\ \hline \hline
0  & 90  &  0.267563 & 2.069346 \\
100& 20  & 0.267581  & 2.069346 \\
25  &-80 & 0.267569  & 2.069346\\
150 &55   & 0.267586 &  2.069346\\
30  &-45 &  0.267568 & 2.069346 \\
\hline \hline
\end{tabular}
\end{center}
\caption{The statistics $\bmath{S_D^p}$ \& $\bmath{Z_c^p}$ for
different choices of arbitrary poles $(\alpha_p,\delta_p)$ with
$n_v = 20$ nearest neighbours.}
\label{invariance}
\end{table}

\section{Results}
\begin{figure}[]
\centering
\includegraphics[]{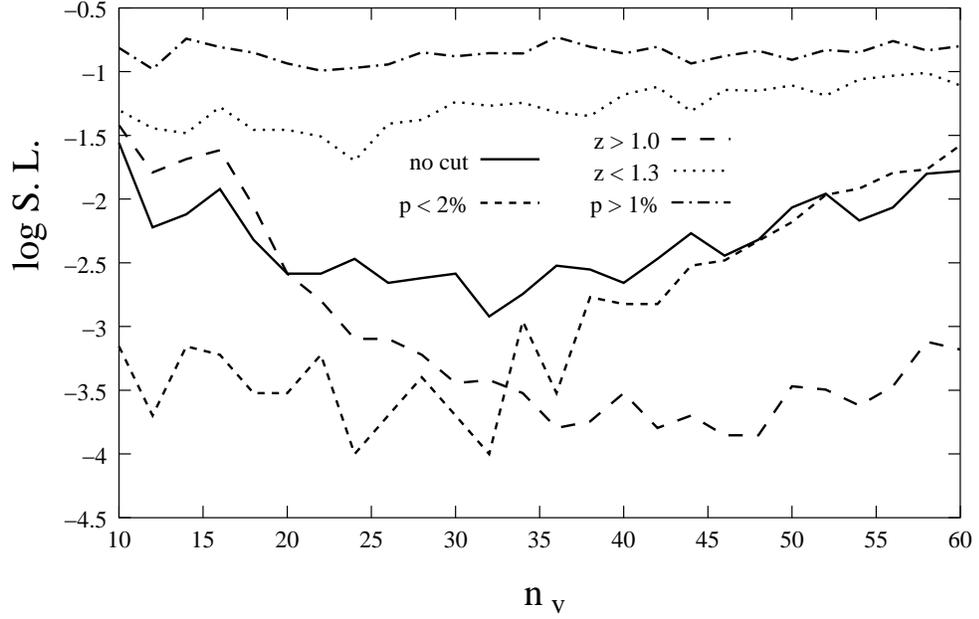}
\caption{ The logarithmic significance level,
$\log(S.L.)$, as a function of the number of 
nearest neighbours $n_v$ using the statistic $S_D^p$. The
nearest neighbours are obtained by taking into account the
radial distance of the source and hence this tests
for redshift dependent alignment. 
The black curve corresponds to the entire data set. The short dashed,
dash-dotted, long dashed and dotted curves correspond to the cuts $p\le 2$\%,
$p\ge 1$\%, $z\ge 1.0$ and $z\le 1.3$ respectively.}
\label{stat1}
\end{figure}

\begin{figure}[]
\centering
\includegraphics[]{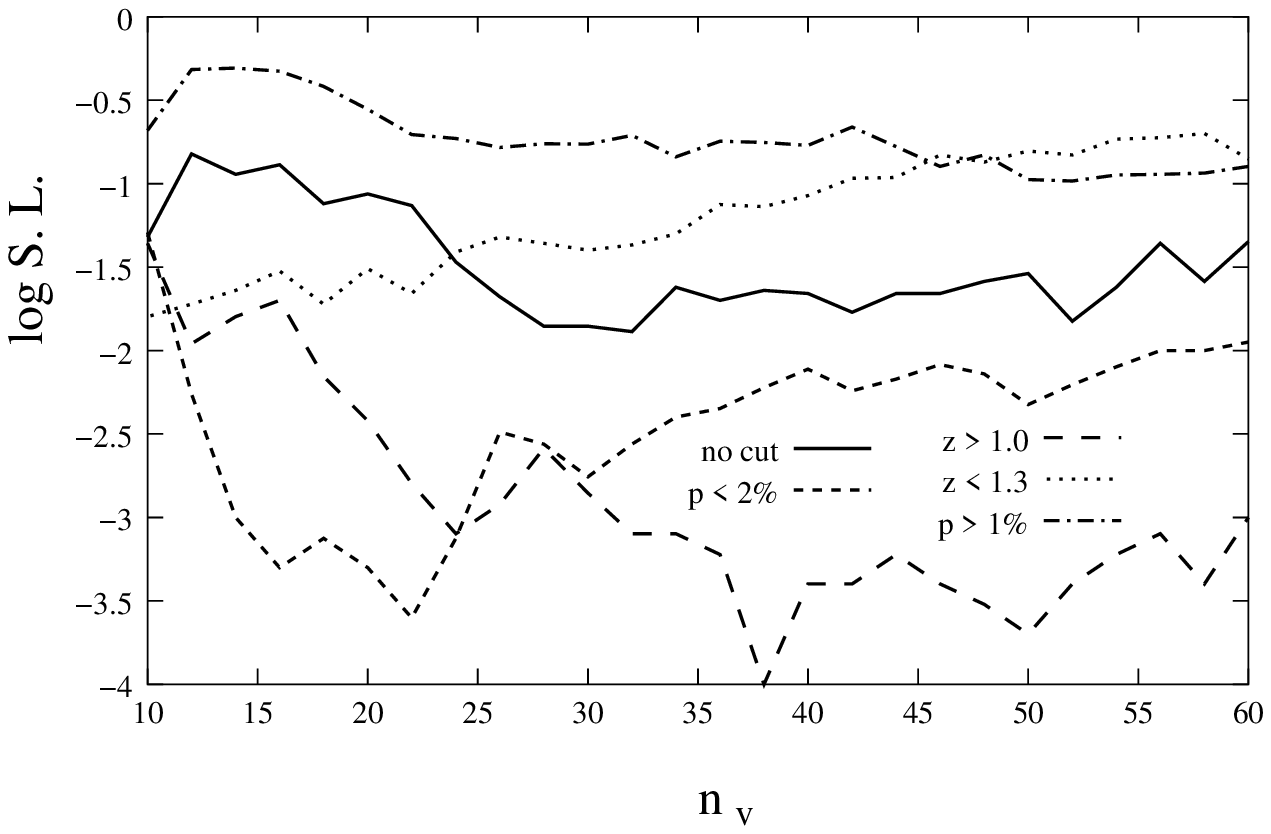}
\caption{ The logarithmic significance level,
$\log(S.L.)$, as a function of the number of 
nearest neighbours $n_v$ using the statistic $S_D^p$. 
The nearest neighbours are obtained without taking into account the
radial distance of the source and hence this tests
for redshift independent alignment. 
The black curve corresponds to the entire data set. The short dashed,
dash-dotted, long dashed and dotted curves correspond to the cuts $p\le 2$\%,
$p\ge 1$\%, $z\ge 1.0$ and $z\le 1.3$ respectively.}
\label{stat2}
\end{figure}

\begin{figure}[]
\centering
\includegraphics[]{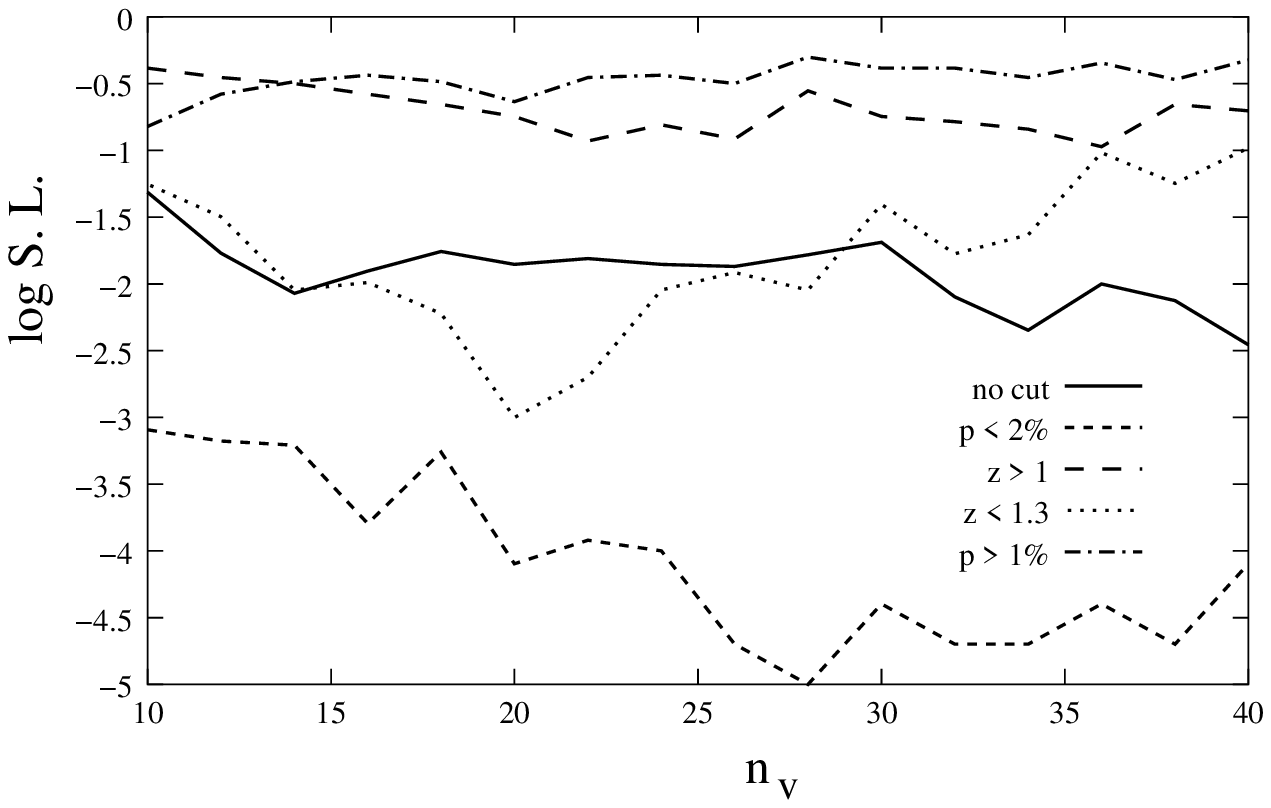}
\caption{ The logarithmic significance level,
$\log(S.L.)$, as a function of the number of 
nearest neighbours $n_v$ using the statistic $Z_c^p$. The
nearest neighbours are obtained by taking into account the
radial distance of the source and hence this tests
for redshift dependent alignment. 
The black curve corresponds to the entire data set. The short dashed,
dash-dotted, long dashed and dotted curves correspond to the cuts $p\le 2$\%,
$p\ge 1$\%, $z\ge 1.0$ and $z\le 1.3$ respectively.}
\label{stat3}
\end{figure}

\begin{figure}[]
\centering
\includegraphics[]{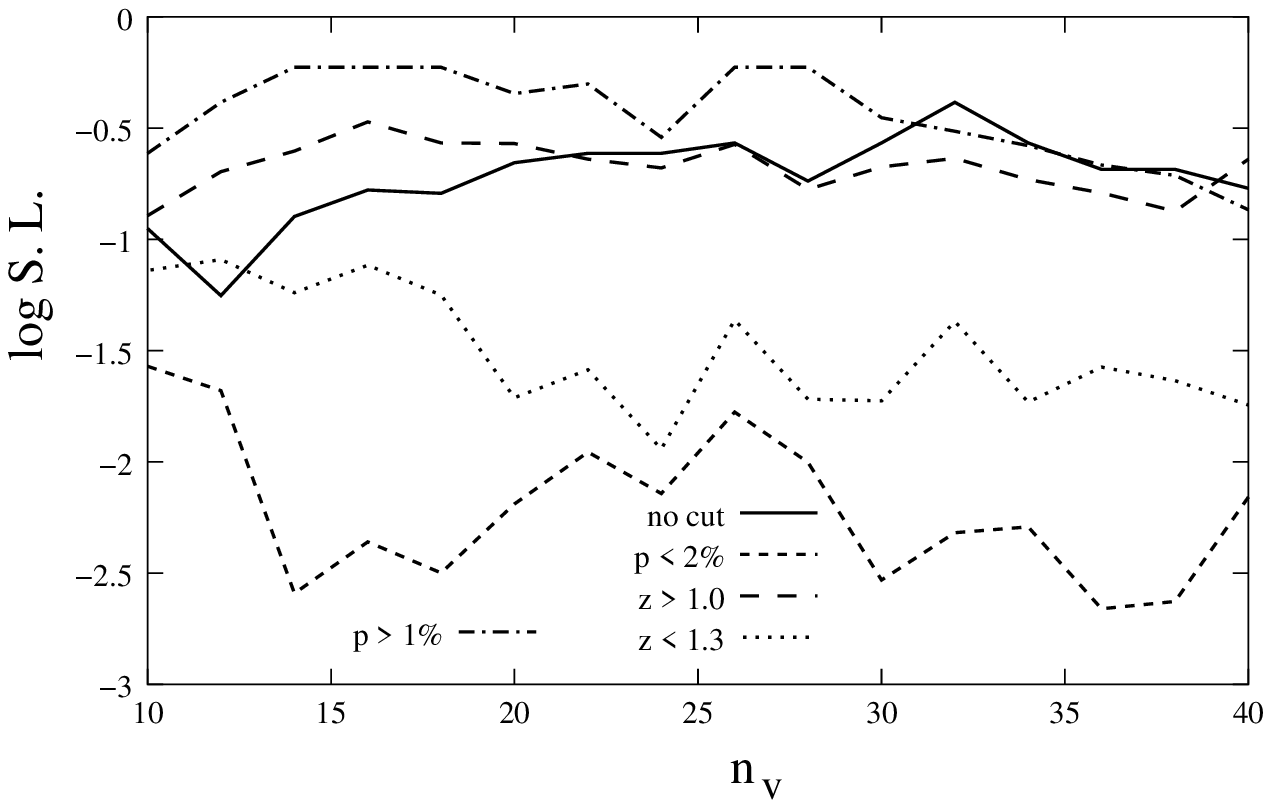}
\caption{ The logarithmic significance level,
$\log(S.L.)$, as a function of the number of 
nearest neighbours $n_v$ using the statistic $Z_c^p$. 
The nearest neighbours are obtained without taking into account the
radial distance of the source and hence this tests
for redshift independent alignment. 
The black curve corresponds to the entire data set. The short dashed,
dash-dotted, long dashed and dotted curves correspond to the cuts $p\le 2$\%,
$p\ge 1$\%, $z\ge 1.0$ and $z\le 1.3$ respectively.}
\label{stat4}
\end{figure}

The significance levels (S. L.) for the $S_D^p$ and $Z_c$ statistic are 
given in Figures \ref{stat1}-\ref{stat4}. 
We plotted the logarithmic significance level  
of both the statistical tests $Z_c^p$ and $S_D^p$ as a function of  
$n_v$, the number of nearest neighbours. The S. L. was computed by 
numerical simulations which compares the data statistic with
that of a large number $N_S$ of random samples.
For most cases the choice $N_S=1000$ was found to be sufficient. 
In several cases it was found necessary to increase 
$N_S$ to 10000 or larger in order 
to get a reasonable estimate of S.L.. 
In computing $Z_c^p$ we used the matrix $D_{ij}^p$ given in Eqn. 
(\ref{dmatrix}). The matrix $D_{ij}^{p\prime}$,
Eqn. (\ref{dmatrixprime}), gave similar results. The results are obtained
both by using the radial  distance $r(z)$ given in Eqn. (\ref{dis}) as
well as by setting $r(z)=1$, which tests for a redshift independent alignment.
We also try out several cuts based on the redshift $z$ and the degree of 
polarization  $p$. In Figs. \ref{stat1},\ref{stat2} we show the 
significance level using 
the $S_D^p$ statistic for redshift dependent and redshift independent alignment
respectively. In Figs. \ref{stat3},\ref{stat4} 
we show the signifigance level using 
the $Z_c^p$ statistic for redshift dependent and redshift independent alignment
respectively.

For an entire data set of 213 points we find that the $\log(S.L.)<-2.5$
over a wide range of $n_v$ values using the $S_D^p$ test for
redshift dependent alignment. The minimum $\log(S.L.)= -2.92$ for
$n_v=32$. Hence we find that there is evidence for large scale 
alignment as observed by \huts (1998). The $Z_c^p$ statistic
(Fig. \ref{stat3}) also shows alignment with $log(S.L.)<-2$ for
$n_v\ge 32$. The redshift independent test (Fig. \ref{stat2} and
\ref{stat4}) does not show any evidence of alignment for the complete
sample. 
We also examined several cuts on the data sample in order to
determine whether the alignment arises from large or small degrees of
polarization and redshifts. We find that low polarizations $p\le 2$\%
show a very significant redshift dependent alignment as seen in Figures
\ref{stat1} and \ref{stat3}. The number of data points with this
cut is 146. The $\log(S. L.) = -4.0$ for 
$n_v=24$ and 32 using the $S_D^p$ statistic and reaches $-5.0$ for $n_v=28$
with the $Z_c^p$ statistic. The data with large polarization,
$p\ge 1$\%, however, shows
no evidence of alignment with any choice of the statistic. 
The number of data points in this
case is 147. The fact that low polarizations are significantly aligned
suggests that the effect may arise due to propagation. 

We next compute the mean distance at which the alignment effect is 
most pronounced by directly testing for alignment over a certain
distance, rather than using the number of nearest neighbours. In
this case we test for alignment of a particular object with all 
the objects within a fixed distance of this object. Using the 
$S_D^p$ statistic with $p\le 2$\% we find that the S.L. takes its minimum
value at $r(z) \approx 0.22 (2c/H_0)$, which corresponds to a distance scale
of the order of a Gpc.  

\begin{figure}[]
\centering
\includegraphics[]{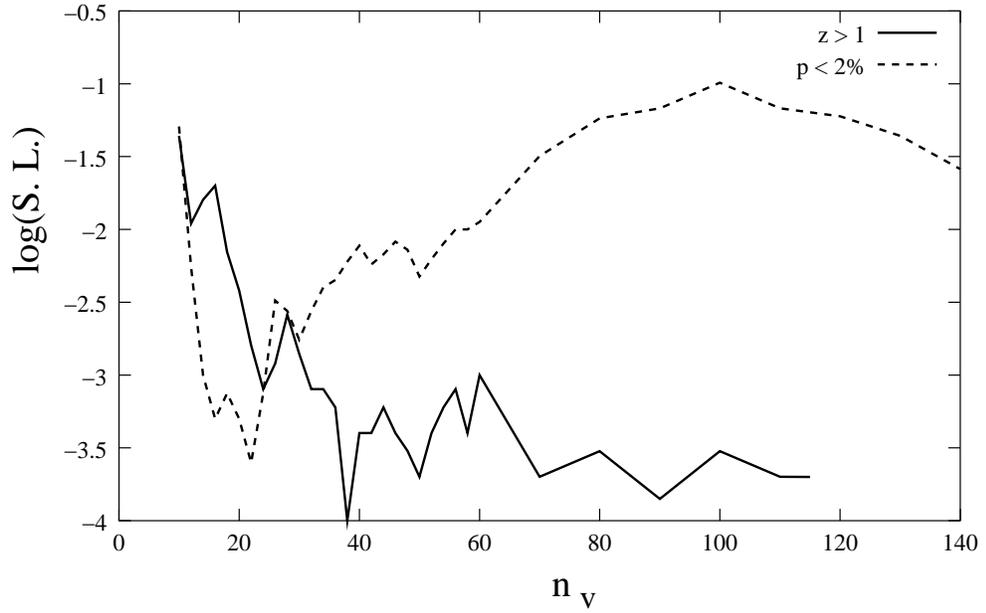}
\caption{ The logarithmic significance level,
$\log(S.L.)$, as a function of the number of 
nearest neighbours $n_v$ using the statistic $S_D^p$ for the
cuts $z\ge 1$ (solid curve) and $p\le 2$\% (dashed curve). The
nearest neighbours are obtained without taking into account the
radial distance of the source and hence this tests
for redshift independent alignment. Results are shown for 
very large values of $n_v$ and this tests for alignment over
very large distances. The total number of points in the sets
$z\ge 1$ and $p\le 2$\% are 115 and 146 respectively.} 
\label{stat5}
\end{figure}

We next examine the
cuts on redshifts. In Fig. \ref{stat1} we find that the sources at
large redshifts $z\ge 1$ show a very strong alignment
with the $S_D^p$ statistic for number of nearest neighbours
larger than about 35. 
The number of objects in the data sample $z\ge 1$ are 115. For comparison
we also consider a low redshift data sample, where the cut $z\le 1.3$ is
chosen such that it also contains approximately the same number of
objects as the $z\ge 1$ sample. The $z\le 1.3$ infact also contains 
115 objects. 
From Figs. \ref{stat1} and \ref{stat2} we find that the large redshift
sample shows a strong alignment independent of whether we take the
radial distance of the object into account or set it equal to unity. 
Hence the
alignment of objects at large redshifts is not necessarily redshift dependent.
The low redshift objects, however, do not show a strong alignment with
this statistic. The results for the $Z_c^p$ statistic (Figs. \ref{stat3}
and \ref{stat4}), however, show
no alignment for large redshift sources and a weak alignment for low
redshifts. 
The fact that the $S_D^p$ statistic shows a strong signal
for alignment of large redshift objects and the $Z_c^p$ shows no signal
is easily understood. We find that the alignment of the large redshift
sources occurs predominantly for relatively large number of nearest 
neighbours, and hence for large distances. The $Z_c^p$ statistic 
dominantly tests only for local alignment. 
The $S_D^p$ can test both local and large scale alignment. The large
redshift points seem to be aligned over very large distances. This is
shown more clearly in Fig. \ref{stat5}, which shows the $\log(S. L.)$
for very large values of $n_v$. The results for $p\le 2$\% cut
are also shown for comparison. 
It is clear from this figure that 
the large redshift points show alignment over the entire sample.
Hence we find that besides the redshift dependent alignment, which
happens primarily for the objects with low polarization, the 
entire set of large redshift objects are aligned with one
another. This effect is different from the redshift 
dependent alignment and was not noticed by \huts (1998).
We emphasize that the redshift dependent alignment is seen
dominantly for low polarizations as shown in fig. \ref{stat1}.
The large redshift objects, however, show an alignment over the
entire sky.

\begin{figure}[]
\centering
\includegraphics[]{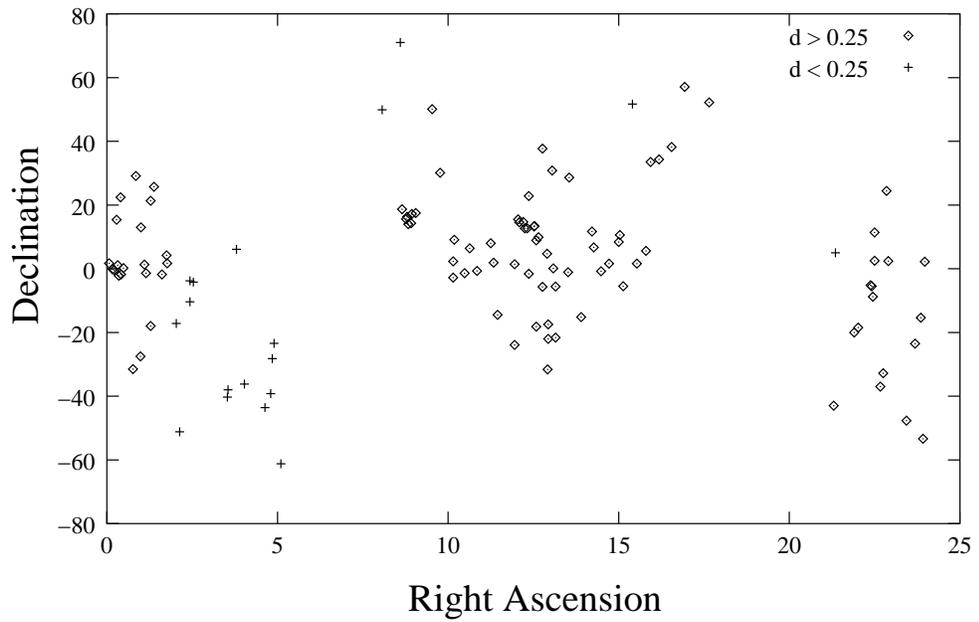}
\caption{Scatter plot of the objects which show significant alignment 
$d_i>0.25$ (dots) and those which do not show alignment $d_i<0.25$ (pluses). 
Here $d_i$ is
a measure of the dispersion as defined in Eq. \ref{dispersion}. The figure
shows the data sample with the cut $z\ge 1$ with the number of nearest
neighbours $n_v = 38$.} 
\label{scatter_zg1}
\end{figure}

In order to understand this very large scale alignment at large
redshifts, $z\ge 1$, we make a scatter plot of the objects which
show a significant alignment along with those which do not.
In this study we ignore the redshift dependence of the objects
since the $z\ge 1$ set shows significant effect irrespective of
whether we include or ignore the redshift dependence. 
From fig. \ref{stat5} we find that the S.L. is minimum at 
$n_v=38$ and hence we choose this value of $n_v$ for our study.
The resulting scatter plot is shown in fig. \ref{scatter_zg1} 
where the pluses and dots refer to the objects for which the dispersion
measure $d_i<0.25$ and $d_i\ge 0.25$ respectively. We point
out that the statistic $S_D^p=0.307$ in this case. We choose $d_i=0.25$
as the cutoff in this figure since it shows the boundary between
the aligned and non-aligned sources clearly. A choice of $d_i=0.3$
as the cutoff also leads to a similar scatter graph, but with a larger
overlap between the aligned and non-aligned regions.
Furthermore the figure does not change too much if we choose a 
larger value of $n_v$. 
It is clear from figure \ref{scatter_zg1} that 
a large contribution to statistic is obtained from what is called
the A1 region by \huts (1998). This region is centered roughly at the
Virgo supercluster and is delimited in redshift such that $1\le z\le 2.3$. 
However as can be seen from Fig. \ref{scatter_zg1},
other regions also show significant alignment.

In order to 
understand the nature of this very large scale alignment we first observe that
the objects lying within the  
coordinate interval $10\le RA\le 16$ and $-30\le Dec\le 30$ are aligned
with one another.
Similarly the objects in the region $RA\ge 22$, $RA\le 2$ and 
$-40\le Dec\le 30$ are also aligned over this entire patch. 
Hence we find that the data splits into two large patches such that 
in each patch most of the objects show alignment with one another.
The mean polarization angles,
i.e. the mean over the $n_v=38$ nearest neighbours,
in the interval $10\le RA\le 16$ are centered at the value of approximately
2.95 radians. In the region $RA\ge 22$, $RA\le 2$, the means are centered
around the value 2.35 radians. A parallel transport from $RA=0,\ Dec=-5.0$
to $RA=13,\ Dec=0$ leads to a shift in the polarization angle by
0.64 radians, which when added to 2.35 leads to an angle very close to
2.95. Hence we find that even these two widely separated regions
are correlated. This explains the significant alignment observed 
in the entire data sample corresponding to $z\ge 1.0$ as shown in 
fig. \ref{stat5}.

\section{Physical Explanation}
The observation of such a large scale alignment is quite surprising
and is not easily explained in terms of conventional astrophysics.
QSOs at such large distances from one another are unlikely to be
correlated with one another and hence the effect most likely arises
due to propagation. The fact that the alignment is most dominant
in the data sample with small polarizations further supports this
hypothesis. Polarizations can show alignment with one another
if they arise dominantly due to extinction. For example the  
galactic magnetic fields result in a large scale alignment of 
the dust particles, which preferentially attenuate a particular
polarization component of the electromagnetic waves. Hence if the
intrinsic source polarization is negligible, this will give rise to 
polarizations which are aligned over large distances. Galactic extinction,
however, is unable to explain the redshift dependence of the effect.
Furthermore it cannot explain why the large redshift points are 
correlated over such large distances, whereas the low redshift 
data shows no such correlation. We next examine the possibility that 
the alignment arises due to supercluster extinction. We assume that
the magnitude of the supercluster extinction is large enough to cause
such an alignment. The redshift dependent alignment
seen in the low polarization sample $(p\le 2\%)$ can then be explained
if we assume the presence of a few very large superclusters of 
distance scale of order 1 Gpc. The cosmological scale alignment seen in the 
large redshift sample is, however, not easily explained by this mechanism 
since we do not expect the presence of a supercluster which covers almost
the entire celestial sphere at redshift of 1. 
Hence the observations are not completely explained in terms
of galactic or supergalactic extinction.

Another physical phenomenon that can potentially explain the observations
is the pseudoscalar-photon mixing (\huts \& Lamy 2001; Jain, Panda \& 
Sarala 2002). A light pseudoscalar
particle is predicted by many extensions of the standard model of
particle physics (for e.g. see
Peccei \& Quinn 1977; Mann \& Moffat 1981; Sachs 1982;
Will 1989; Ahluwalia \& Goldman 1993; Ralston 1995). 
Its mixing with photons and its 
astrophysical consequences have also been studied by 
several authors (for e.g. see Sikivie 1983;  
Maiani, Petronzio \& Zavattini 1986; Harari \& Sikivie 1992; Das, 
Jain \& Mukherjee 2001).
Its coupling to photons is bounded by observations
of SN87A to be $g< 10^{-11}$ GeV$^{-1}$ (Brockway, Carlson \& Raffelt 1996; 
Grifols, Masso \& Toldra 1996; Raffelt 1999;  Rosenberg \& van Bibber 2000).
Such a particle has also be invoked to explain the observed dimming
of supernovae at large redshifts even if the expansion rate
of the universe is not accelerating (Csaki, Kaloper \& Terning 2002).
The pseudoscalar particle 
decays into photons in the presence of background magnetic field $\vec B$,
such that the photon produced is polarized parallel to the transverse
component of $\vec B$, which we denote as $\vec B_T$. 
Hence the electromagnetic wave produced by this 
mechanism is polarized parallel to $\vec B_T$.
Similarly a photon polarized parallel to
$\vec B$ can decay into the pseudoscalar during propagation, which leads
to a wave polarized perpendicular to $\vec B_T$. 

The existence of a hypothetical pseudoscalar can explain the
observed alignment as follows. We first assume that the redshift
dependent effect seen in low polarization objects is explained 
by the presence of a few superclusters of length scales of order 1 Gpc.
As the electromagnetic waves pass through these superclusters the
polarizations get aligned 
either due to extinction or due to pseudoscalar-photon
mixing. We further assume that the large redshift objects, 
which show alignment over the entire
sky, emit a large flux of the pseudoscalar particles whereas the
flux emitted by low redshift objects is relatively small.
This is reasonable since the large redshift 
objects are in general very active and
have high temperatures. As these QSOs evolve they are less active and
the pseudoscalar flux becomes negligible, which explains why the low
redshift objects behave differently. As the pseudoscalars propagate 
through our local supercluster they decay into photons which are
polarized parallel to the supercluster magnetic field. 
As discussed by Jain et al. (2002) the decay probability of such a 
pseudoscalar is of order unity with the current limits on the
pseudoscalar-photon coupling. In obtaining this estimate we took the
Virgo supercluster parameters for the magnetic field $(B\approx 1\mu G)$
(Vallee 1990) and plasma density $(n_e\approx 10^{-6}$ cm$^{-3})$. 
We propose that a weaker magnetic field, $B\approx 0.1 \mu G$, might
be associated with the entire supercluster, and may be responsible for the
observed alignment of large redshift objects over very large 
angular separations. We point out that the decay probability of pseudoscalars
$\phi$ into photons $\gamma$, $P_{\phi\rightarrow\gamma}\sim B^2/n_e^2$
(Carlson \& Garretson 1994; Jain et al. 2002)
and hence even if the magnetic field is an order of magnitude smaller than
that observed in the Virgo supercluster, $P_{\phi\rightarrow\gamma}$ can
be large as long as $n_e$ also decreases proportionately.
Hence this
phenomenon explains the alignment seen in this sample over very large
angles. 

It is clearly very important to further test the proposal that the
alignment effect is due to pseudoscalar-photon mixing. 
Jain et al. (2002) have computed all the Stokes parameters of
the electromagnetic wave in the presence of such a pseudoscalar 
particle. They find that the spectral dependence 
of the linear polarization, the circular polarization
and the orientation angle of the linear polarization are all closely
correlated with one another due to this mixing. This correlation
can be studied by making further observations, which can establish
or rule out this explanation.

In our discussion we have assumed an isotropic and homogeneous universe.
We have not considered the possibility of a cosmological scale
magnetic field or a Lorentz violating interaction (Nodland \& Ralston 1997;
Carroll, Field \& Jackiw 1990), 
which might also explain these observations. A large scale dipole
anisotropy is also observed in the radio polarizations from distant
AGNs (Jain \& Ralston 1999). This anisotropy was found to be independent of
radial distances and might be explained in terms of some local effect.
The axis of this dipole anisotropy is found
to be pointing approximately opposite to the center of our local 
supercluster. Since this is a dipole anisotropy the effect on polarizations
is large both in the direction and opposite to the axis.
Hence the radio anisotropy displays an intriguing relationship to the
large scale alignment seen in optical polarizations in the direction
of the supercluster center. This might be a hint for a common origin
of these two effects. The fundamental
origin of the radio anisotropy is not known. Jain et al. (2002)
pointed out that the existence of a light pseudoscalar
can also explain this effect if the AGNs emit a large flux of pseudoscalars
at radio frequencies. Although this explanation is disfavored due to
the very large pseudoscalar flux required from the distant AGNs, 
it does explain the 
correlation between the
radio dipole axis and the observed large scale alignment in optical
polarizations. 

\section{Conclusions}
In conclusion we have developed coordinate invariant  
statistical procedures in order to test for the
large scale alignment of optical polarizations.
We applied these tests to a data sample of QSOs compiled by 
\huts (1998) and \huts \& Lamy (2001) and find that the polarizations show
statistically significant alignment over very large distances.
We find that the alignment is redshift dependent and is seen
dominantly for the data sample with low polarizations $(p\le 2\%)$. We also
find that the large redshift, $z\ge 1$, sample shows a very large
scale alignment. Infact the polarizations contained in 
almost the entire data sample at $z\ge 1$ seem to be correlated
with one another. We find that galactic or supercluster extinction
is unlikely to provide an explanation for these observations. We also
argue that the existence of a hypothetical pseudoscalar particle might
provide an explanation for the alignment effect.
 
\bigskip
\noindent
{\bf Acknowledgements:} We thank Gopal Krishna, Rajaram Nityananda,
D. J. Saikia and Tarun Souradeep for useful discussions.

\bigskip
\noindent{\bf References}

\begin{itemize}
\item[] Ahluwalia D. V.,  Goldman T., 1993, Mod. Phys. Lett., A 28, 2623 
\item[] Bietenholz M. F., 1986, AJ, 91, 1249 
\item[] Brockway J. W., Carlson E. D., Raffelt G., 1996,
Phys. Lett., B 383, 439 (astro-ph/9605197)
\item[] Carlson E. D., Garretson W. D., 1994, Phys. Lett., B 336, 431 
\item[] Carroll S. M., Field G. B., Jackiw R., 1990, Phys. Rev., D 41, 1231 
\item[] Csaki C., Kaloper N., Terning J., 2002, Phys. Rev. Lett.,
88, 161302, (hep-ph/0111311)
 \item[] Das P., Jain P., Mukherjee S., 2001,
  Int. Jour. of Mod. Phys., A 16, 4011 (hep-ph/0011279).
\item[] Grifols J. A., Masso E., Toldra R., 1996, Phys. Rev. Lett., 
77, 2372
\item[] Harari D., Sikivie P., 1992, Phys. Lett., B 289, 67 
\item[] \huts  D., 1998, A \& A, 332, 410
\item[] \huts D., Lamy H., 2001, A \& A, 367, 381 
\item[] Jain P., Panda S., Sarala S., 2002,
Phys. Rev., D 66, 085007 (hep-ph/0206046)
\item[] Jain P., Ralston J. P., 1999, Mod. Phys. Lett., A 14, 417 
\item[] Maiani L., Petronzio R., Zavattini E., 1986,
Phys. Lett., B 175, 359
\item[] Mann R. B., Moffat J. W., 1981, Can. J. Phys., 59, 1730 
\item[] Nodland B., Ralston J. P., 1997, Phys. Rev. Lett., 78, 3043 
\item[] Peccei R. D., Quinn H., 1977, Phys. Rev., D 16, 1791
\item[] Raffelt G., 1999, Ann. Rev. Nucl. Part. Sci.,
49, 163 (hep-ph/9903472)
\item[] Ralston J. P., 1995, Phys. Rev., D 51, 2018 
\item[] Rosenberg L. J., van Bibber K. A., 2000,
Phys. Rep., 325, 1 
\item[] Sachs M., 1982, General Relativity and Matter, Reidel,
\item[] Sikivie P., 1983, Phys. Rev. Lett.,  51, 1415
\item[] Vallee J. P., 1990, AJ, 99, 459
\item[] Will C. M., 1989, Phys. Rev. Lett., 62, 369 

\end{itemize}

\end{document}